\documentclass[11pt]{article}
\usepackage{SIGACTNews}
\usepackage{times}
\usepackage{indentfirst}
\usepackage{graphicx}
\usepackage{amsthm}
\usepackage[margin=1in]{geometry}
\DeclareGraphicsExtensions{.pdf,.png,.jpg}
\newtheorem{theorem}{Theorem}
\title{Using Quadrilaterals to Compute the Shortest Path \\ \[Extended Abstract\]}
\author{
Newton Campbell Jr.\\
       Nova Southeastern University\\
       nc607@nova.edu
\footnote{This research was conducted at Nova Southeastern University Graduate School of Computer and Information Sciences in partial fulfillment for the requirements for the degree of the Doctor of Philosophy (Ph.D.) in Computer Science. Tuition assistance for the program was provided by the student's employer, Raytheon BBN Technologies.}
}
\begin{document}
\SIGACTmaketitle
\begin{abstract}
We introduce a new heuristic for the A* algorithm that references a data structure of size $\theta(|\textbf{\emph{L}}|^{2} + |V|)$, where $\textbf{\emph{L}}$ represents a set of strategically chosen landmark vertices and $V$ the set of vertices in the graph. This heuristic's benefits are permitted by a new approach for computing landmark-based lower bounds, in which each landmark stores only the distances between it and vertices within its graph partition. During search queries, a geometric inequality based on distance information for multiple landmarks is used to establish a lower bound for the search. In comparison to previous landmark-based algorithms, this process significantly reduces the amount of preprocessed distance information that needs to be stored (typically $\theta(|\textbf{\emph{L}}|\cdot|V|)$) while also granting a constant computational cost for each vertex visited during a shortest path query. Further, for graphs with non-overlapping partitions, preprocessing this data structure requires time complexity equivalent to one Dijkstra's shortest path tree computation on the graph.
\par
We characterize the behavior of this new heuristic based on a dual landmark configuration that leverages quadrilateral inequalities to identify the lower bound for shortest path. Using this approach, we demonstrate both the utility and detriments of using polygon inequalities aside from the triangle inequality to establish lower bounds for shortest path queries. While this new heuristic does not dominate previous heuristics based on triangle inequalities, the inverse is true, as well. Further, we demonstrate that an A* heuristic function does not necessarily outperform another heuristic that it dominates. In comparison to other landmark methods, the new heuristic maintains a larger average search space while commonly decreasing the number of computed arithmetic operations. The new heuristic can significantly outperform previous methods, particularly in graphs with larger path lengths. The characterization of the use of these inequalities for bounding offers insight into its applications in other theoretical spaces.
\end{abstract}
\pagebreak
\section{Introduction}
	Since Euler's early analysis of the map of Königsberg \cite{Euler1736}, a large body of work has been devoted to pathfinding problems such as the point-to-point shortest path problem. As researchers find more use for graph theory in the storage, retrieval, and analysis of big data, extremely fast solutions to problems such as the shortest path problem are in great demand. Not even Dijkstra's or the A* algorithm can solve the problem for massive datasets without an exponential increase in their requirements for computational time and space. Consequently, modern research focuses on computing and storing a data structure derived from the graph structure prior to allowing it to be queried for shortest path. This data structure is used to guide, narrow, or inform the search such that arbitrary queries can be performed significantly faster on graphs that represent huge data corpuses.  Modern approaches typically exploit mathematical estimations \cite{Delling2007,Delling2009,GoldbergHarrelson2005,Maue2010}, large-scale storage \cite{Duan2009,Goldman1998,Thorup2001,Sankaranarayanan2010}, artificial intelligence algorithms \cite{Awasthi2005,Yussof2009,Zakzouk2010,Zongyan2012}, and combinations of preprocessing algorithms \cite{Sanders2007}.
\par
We introduce an approach for shortest path estimation that was derived from ALT\cite{GoldbergHarrelson2005}, a class of algorithms that leverage \underline{A}* search, a set of strategically chosen vertices called \underline{l}andmarks, and the \underline{t}riangle inequality to obtain lower bounds for shortest path queries. ALT works as follows: prior to search, the distances between a chosen set of vertices in a graph and all other vertices are computed and stored in a data structure. For a set of landmarks $\textbf{\emph{L}}$, graph vertex set $|V|$, and graph edge set $|E|$, this data structure is of size $\theta(|\textbf{\emph{L}}|\cdot|V|)$ and is computed in $\theta(|\textbf{\emph{L}}| \cdot (|E|+|V|  \log{|V|}))$ time. During search queries, source and target information contained in this data structure is leveraged to identify a triangle in the graph, allowing the triangle inequality to be established as a lower bound for the shortest path. Our approach, known as ALP, is a class of algorithms that leverage \underline{A}* search, \underline{l}andmarks, and (other) \underline{p}olygon inequalities. In this paper, demonstrate that ALP leverages a data structure of size $\theta(|\textbf{\emph{L}}|^{2} + |V|)$ as opposed to ALT's previous $\theta(|\textbf{\emph{L}}|\cdot|V|)$. This data structure is permitted by a new embedding process for identifying distances only within a graph partition. The scenarios and configurations in which each heuristic outperforms the other are enumerated. Finally, we highlight the ability for Dijkstra's algorithm to outperform both ALT and ALP.
\subsection{Previous Work on Shortest Path Preprocessing}
Significant work has been done in preemptively analyzing graphs to store information that can assist in solving the point-to-point shortest path (PPSP) problem. Geometric Containers \cite{Wagner2005} algorithms relies on the concept of edge labeling, where preprocessing attaches a label to each edge in a graph that represents all nodes to which a shortest path starts with the individual edge. This means that for a graph $G = (V, E)$, Geometric Containers maintain a linear space requirement of $\theta(|E|)$. However, the preprocessing step requires a single source shortest path search from every node, requiring $\theta(|V| \cdot (|E|+|V| \log{|V|}))$ preprocessing time. Arc Flags algorithms \cite{Rolf2007} require an input graph that is partitioned such that a flag is computed for each edge within a partition, or region, which indicates whether the edge is on a shortest path to any node in that partition. It is similar to the Geometric Containers in that it considers only the edges whose flag correspond to a specific region. This algorithm realizes a high preprocessing time, as one shortest path search from every border node of a region is required. For $B$ border nodes and $k$ partitions, Arc Flags preprocessing maintains a time complexity of $\theta(B \cdot |E|+|V| \log{|V|})$ with a space requirement of $\theta(k \cdot |E|)$.
\par
The precomputed cluster distances (PCD) algorithm \cite{Maue2006} was designed with the intention of reducing the space requirements of preprocessing algorithms such as ALT \cite{GoldbergHarrelson2005} (detailed in the next section). PCD leverages the distances between graph clusters to compute the heuristic for A*. The algorithm has demonstrated practical performance benefits over the ALT, Arc Flags and Geometric Containers preprocessing algorithms. It is often noted that the space complexity for PCD is $\theta(k^{2}+B)$ compared to ALT's $\theta(|\textbf{\emph{L}}| \cdot |V|)$, where $k$ is equal to the number of graph clusters, $B$ is equal to the number of border nodes for clusters, and $L$ is the set of chosen landmark vertices. However, since the actual clustering information is stored for PCD, as well, the actual space complexity is $\theta(k^{2}+B+|V|)$, as information about which cluster every vertex belongs to needs to be maintained by a data structure in order to be referenced. PCD still boasts a higher experimental average speedup for PPSP queries compared to ALT and has a smaller preprocessing time complexity.
\par
Reach-based pruning is another method for speeding up shortest-path queries such as Dijkstra's algorithm. Reach is a centrality measure that identifies how central a vertex is on a shortest path \cite{Gutman2004}. The latest methods for preprocessing the reach for all nodes requires $O(|V|^{2} \log{V})$ time and $O(|V|)$ storage. Reach-based pruning is often combined with ALT to perform REAL, a method that stores landmark distances of only high reach\cite{GoldbergKaplan2006}.
\subsection{Outline}
We begin the technical part of this paper in Section 2 with a review of basic graph theory, the ALT class of algorithms, and embedding methods for ALT. In Section 3, we describe the ALP approach in detail, with a focus on identifying quadrilaterals in the graph. In Section 4, we discuss a new method of forming a landmark-based preprocessing data structure called distributed embedding. In Section 5, we discuss how different heuristics should be compared for A* to more closely map heuristic metrics to computational performance. Finally, Section 6 describes how the research can be carried forward.
\section{Using Landmarks To Calculate Shortest Path}
\subsection{Graph Notation}
Throughout the rest of this paper, a common set of graph theoretical definitions, concepts, and notations will be used. Let $G = (V,E)$ be an undirected graph, where $V$ is the set of vertices in $G$ and $E \subseteq V \times V$ is the set of edges in $G$, with $n = |V|$ and $m = |E|$. For any edge $e \in E$, let $w(e)$ be the positive real \emph{weight} of $e$. In an \emph{unweighted} graph, for every edge $e \in E, w(e) = 1$. In a \emph{weighted} graph, $w(e)$ is subject to the graph's application. A \emph{finite} graph is one in which $|V|\neq\infty$ and $|E|\neq\infty$. If an edge $e \in E$ connects two vertices $v_i,v_j \in V$, $v_i$ is called the \emph{neighbor} of $v_j$ and $v_j$ the \emph{neighbor} of $v_i$. The vertices $v_i$ and $v_j$ are also said to be \emph{adjacent} to each other and \emph{incident} to their shared edge $e$. A graph $H = (V(H), E(H))$ is a \emph{subgraph} of $G$ if $V(H) \in V$ and $E(H) \in E$, with edges of $E(H)$ incident to only the vertices in $V(H)$. A \emph{spanning subgraph} $H$ of $G$ is a subgraph in which $V(H) = V$. An \emph{induced subgraph} $H$ of $G$ is a graph whose set of vertices are a subset of the vertices of $G$ and whose set of edges have only endpoints in $V(H)$. A graph \emph{cluster}, or \emph{community} is a collection of vertices in a graph such that the vertices assigned to a particular community are similar or connected by some predefined criteria.
\par
A sequence $(v_{0},...,v_{k-1})$, $k \ge 1$, of vertices of $G = (V,E)$ is known as a \emph{path} from $v_0$ to $v_{k-1}$ if there is an edge $(v_i, v_{i+1}) \in E$ for every $0 \le i < k$. A path is denoted as $P(v_0,v_{k-1}) = ‹v_0,...,v_{k-1}›$. A path $P$ is also a subgraph of $G$. The \emph{path length} of $P$ is the number of edges (i.e., $k-1$) on the path $P(v_0,v_{k-1})$, denoted as $d(v_0,v_{k-1})$ or $d(P)$. If a path is the shortest path, its path length is referred to as the \emph{distance} from $v_0$ to $v_{k-1}$. The \emph{path weight} of $P$ is the sum of the weights of the path edges, denoted as $w(P)$ or $w(v_0,v_{k-1})$. If, for every pair of vertices $v_i, v_j \in V$, there exists a path from $v_i$ to $v_j$, the graph is called \emph{connected}. An acyclic, connected, spanning subgraph of $G$ is called a \emph{spanning tree} of $G$.  In this paper, the algorithms focus on undirected, unweighted, finite, connected graphs.
\subsection{ALT (A*, Landmarks, and Triangle Inequalities)}
In ALT, a shortest path query uses a computed distance estimate, derived from the triangle inequality, to guide the search. Let $\textbf{\textit{L}} \in V$ be the set of landmarks with distance $d(v, l_i)$ stored for all vertices $v \in V$ and any landmark $l_i \in \textbf{\textit{L}}, 1 \le i \le | \textbf{\textit{L}}|$. Using the shortest path distances for graph $G = (V,E) $, this inequality yields two important equations for any three vertices $s,t,l_i \in V$:
\begin{equation}
    \label{triangle_inequality_upper_bound}
    d(s,t) \le d(s,l_i )+ d(l_i,t)
\end{equation}
\begin{equation}
    \label{triangle_inequality_lower_bound}
    d(s,t) \ge |d(s,l_i )- d(l_i,t)|
\end{equation}
Based on these inequalities, ALT works as follows: In a preprocessing step, the Dijkstra's shortest path tree (SPT) algorithm is used to compute and store the distances to each landmark in $\textbf{\textit{L}}$ from all other vertices in $V$. Then, during shortest path distance queries, the triangle inequality is used as follows: let $\pi_t^L (v)$ denote the heuristic function based on landmarks used for the A* algorithm. The following equation represents the ALT heuristic function when visiting vertex $v \in V$ on the way to a target vertex $t$:
\begin{equation}
    \label{altheuristic}
    \pi_{t}^{L} (v) = \max_{l_i \in L}{|d(v,l_i )- d(t,l_i )|}
\end{equation}
\subsection{Embedding Methods}
\label{subsec:embeddingmethods}
Briefly, we review the most common \emph{embedding methods}, or \emph{landmark selection techniques}, for ALT. \emph{Random} landmark selection \cite{GoldbergHarrelson2005} identifies and tests $k$ vertices at random to serve as landmarks (where $k$ is typically a parameter). In terms of lower bounds, random landmark selection demonstrate better performance than any of the following methods of landmark selection \cite{GoldbergWerneck2005}. \emph{Farthest} landmark selection \cite{GoldbergHarrelson2005} identifies vertices that maximize path weight between each other. A later algorithm for farthest landmark selection was created to maximize path distance instead of path weight \cite{GoldbergWerneck2005}. This biases farthest selection to choose separate, dense regions of the graph to place landmarks in. \emph{Planar} landmark selection \cite{GoldbergHarrelson2005} uses graph layout information to divide a graph into sectors and then executes \emph{farthest} landmark selection. \emph{Avoid} landmark selection, a commonly used and modified landmark selection algorithm, computes the SPT $T_r$, rooted at a random vertex $r \in V$ \cite{GoldbergWerneck2005} and strategically identifies leaves of the tree to serve as landmarks. This approach "avoids" existing landmarks to improve coverage of landmarks over the graph.
\section{A*, Landmarks, and Polygon Inequalities}
\subsection{Quadrilateral-Based Inequalities in Graphs}
ALT computes shortest path trees (SPTs) from a selected set of landmarks and uses the triangle inequality at query time to establish a lower bound for A* \cite{GoldbergHarrelson2005}. Here, we demonstrate that these heuristics can be derived from other geometric inequalities by identifying other types of polygons in a graph and setting the heuristic values for A* equal to the maximum lower bound. The ALT algorithm is the base case for such a hypothesis. The use of two landmarks, as seen in this paper, provides an inductive step for the proof of the hypothesis. To show this, first, we describe how to form a triangle in a graph to establish the triangle inequality as a lower bound. This proof was derived from the  reverse triangle inequality proof for a metric space (detailed in full paper).
\par
For a simple graph $G$ with vertices $A,B,C \in V$, the shortest path distances between each vertex allow the graph to form a metric space. Therefore, for the distances between vertices $A,B,C \in V$, the following triangle inequalities hold:
\begin{equation}
    \label{triangle_inequality_upper_bound_ab}
    d(A,B)\le d(A,C)+ d(C,B)
\end{equation}
\begin{equation}
    \label{triangle_inequality_upper_bound_bc}
    d(B,C)\le d(B,A)+ d(A,C)
\end{equation}
Both of these inequalities apply to the three vertices in $G$. The reverse triangle inequality, which is used as a lower bound for A* in ALT, is derived from these inequalities. ALT uses this reverse triangle inequality to create a heuristic that estimates the distance between vertices $A$ and $C$ by setting vertex $B$ equal to a landmark $l$ such that
\begin{equation}
    \label{triangle_inequality_lower_bound_abl}
    |d(A,l)-d(l,C)| \le d(A,C)
\end{equation}
By computing and storing the values $d(A,l)$ and $d(l,C)$ before performing any PPSP queries, this lower bound is then used as a heuristic to the A* algorithm. Because it is the lower bound, it will never overestimate the distance between vertices $A$ and $C$.
\par
\begin{figure}
\centering
\includegraphics[scale=0.55]{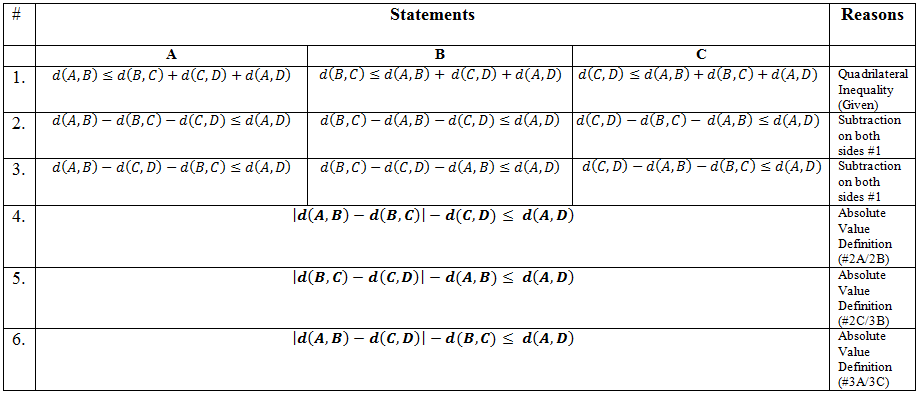}
\caption{Derivation of the Reverse Quadrilateral Inequality in Simple, Connected Graphs}
\label{fig:quadrilateralinequalityproof}
\end{figure}
For a quadrilateral, the lower bound of one of its sides can also be calculated using the other three sides. This reverse quadrilateral inequality can also be used to establish the lower bounds for the shortest path of a graph. For a graph $G$ with vertices $A,B,C,D \in V$, Figure \ref{fig:quadrilateralinequalityproof} demonstrates how we derived the following reverse quadrilateral inequalities based on upper bound estimates:
\begin{equation}
    \label{reverse_quads_1}
    |d(A,B)-d(B,C)|-d(C,D) \le d(A,D)
\end{equation}
\begin{equation}
    \label{reverse_quads_2}
    |d(C,D)-d(A,B)|-d(B,C) \le d(A,D)
\end{equation}
\begin{equation}
    \label{reverse_quads_3}
    |d(B,C)-d(C,D)|-d(A,B) \le d(A,D)
\end{equation}
A potential problem with these inequalities is that they have ability to generate negative lower bound estimates, which is useless for a nonnegative distance metric. For utility, when attempting to estimate the lower bounds of a quadrilateral, other inequalities should be considered such that the highest possible lower bound can be used. In this paper, we use Ptolemy's inequality \cite{Kay2011} to demonstrate an example of this. Ptolemy's inequality can be applied to the graph quadrilateral as follows to yield lower bounds for the distance between $A$ and $D$. We begin with the original inequality:
\begin{equation}
    \label{ptolemys}
    d(A,C)\times d(B,D) \le d(A,B) \times d(C,D) + d(B,C) \times d(A,D)
\end{equation}
Then to estimate the distance between $A$ and $D$, using simple algebra,
\begin{equation}
    \label{ptolemys_alg}
    {{d(A,C)\times d(B,D) - d(A,B) \times d(C,D)} \over {d(B,C)}}  \le d(A,D)	
\end{equation}
In practical cases, information regarding the values of $d(A,C)$ and $d(B,D)$ (the diagonals) may be unknown. Therefore, the distance between can be estimated as follows. First, suppose all the values on the right side of Equation \ref{ptolemys} are known (except, of course, the distance between vertices A and D) and the values on the left side are unknown. Using the reverse triangle inequality, we understand that
\begin{equation}
    \label{ptolemys_alg_rev1}
    1 \le |d(A,B)-d(B,C)| \le  d(A,C)	
\end{equation}
\begin{equation}
    \label{ptolemys_alg_rev2}
   1 \le |d(B,C)-d(C,D)| \le  d(B,D)	
\end{equation}
for $A \ne B \ne C \ne D$. Because they are non-negative, we also know that
\begin{equation}
    \label{ptolemys_alg_nn}
   1 \le |d(A,B)-d(B,C)| \times |d(B,C)-d(C,D)|\\
    \le d(A,C) \times d(B,D)	
\end{equation}
Using these lower bounds, we can rewrite Ptolemy's inequality with respect to the lower bound for the distance between vertices A and D as
\begin{equation}
    \label{ptolemys_heur}
    {{|d(A,B)-d(B,C)| \times |d(B,C)-d(C,D)| - d(A,B) \times d(C,D)} \over {d(B,C)}} \le d(A,D)	
\end{equation}
Because we use Ptolemy's inequality here, this can become a perfect estimate when a cyclic quadrilateral is formed from the four endpoint vertices, $A,B,C,D \in V$.
\par
Because multiple points are used, more inequalities can be generated to estimate distances. The maximum over the set of lower bounds derived by these inequalities can be used to tighten the lower bound for the distance between $A$ and $D$. With that said, two more lower bounds for $A$ and $D$, derived from the triangle inequality, are noted here:
\begin{equation}
    \label{quad_tri_1}
   |d(A,B)-d(B,D)| \le d(A,D)	
\end{equation}
\begin{equation}
    \label{quad_tri_2}
   |d(A,C)-d(C,D)| \le d(A,D)
\end{equation}
\subsection{Dual Landmark Heuristics for A*}
\begin{table*}
\centering
\begin{tabular}{|c|c|}
\hline
1& $\pi_t^{DL}(v,1)=|d(v,l_1 )-d(l_1,l_2 )|-d(l_2,t)$\\    \hline
2& $\pi_t^{DL}(v,2)=|d(v,l_1 )-d(l_2,t)|-d(l_1,l_2 )$\\    \hline
3& $\pi_t^{DL}(v,3)=|d(l_1,l_2 )-d(l_2,t)|-d(v,l_1 )$\\    \hline
4& $\pi_t^{DL}(v,4)=|d(v,l_1 )-d(l_1,t)|$\\   \hline
5& $\pi_t^{DL}(v,5)=|d(v,l_2 )-d(l_2,t)|$\\   \hline
6& $\pi_t^{DL}(v,6)={{|d(v,l_1 )-d(l_1,l_2)| \times |d(l_1,l_2 )-d(l_2,t)|-d(v,l_1) \times d(l_2,t)} \over {d(l_1,l_2)}}$\\   \hline
\end{tabular}
\caption{Dual Landmark Heuristics for ALP}
\label{dual_h_tab}
\end{table*}
Just as with triangle inequalities, the lower bound produced by the quadrilateral inequalities in the previous subsection can be used as heuristics for the A* algorithm. By choosing two landmark vertices to act as endpoints $B$ and $C$, new heuristics are computed as follows: At the visited vertex and target nodes $v,t \in V$ and two valid landmark vertices $l_1,l_2 \in V$ in a graph $G$, let $\pi_t^{DL}(v,i), i \in [1,6]$ denote each new heuristic function for A*. Table \ref{dual_h_tab} lists each heuristic based on the mentioned lower bounds. Each of these are new, admissible heuristics for A* based on polygon inequalities, specifically for quadrilaterals. The following is the optimal dual landmark heuristic now proposed for ALP:
\begin{equation}
    \label{dual_alp}
    \pi_t^{DL}(v)= \max_i{\pi_t^{DL} (v,i)}
\end{equation}
Just as with ALT, the maximum over the set of available lower bounds is used to tighten the lower bounds. The A* algorithm is used with this new heuristic function as input, just as in ALT, with one change. This change involves a process known as \emph{distributed embedding}. The distributed embedding process is further detailed in a later section. In summary, after landmark selection, each vertex in the graph is assigned to a single landmark. These vertices contain distance information for only the landmark to which they are assigned. As a vertex $v$ is visited, if $v$ does not have distance information at its current landmark node, $l_1$, the landmark that does have distance information for $v$ is used to bound the search. For unidirectional A*, the $l_2$ landmark remains the same for the target node, as it is the only one containing distance information for that node. This fact, of course, would change for the bidirectional variant of A*. Note that, when using distributed embedding, $\pi_t^{DL}(v,4)$ and $\pi_t^{DL}(v,5)$  can only be used when both the visited node $v$ and target node $t$ share the same landmark. Otherwise, the information needed for this heuristic cannot be computed. If the source and target vertex share the same landmark (i.e., $l_1=l_2$), then the ALP heuristic is reduced to the ALT heuristic (i.e., $l_1=l_2=l$) as follows:
\begin{equation}
    \label{dual_alp_1}
    \pi_t^{DL}(v,1) = |d(v,l)-d(l,l)|-d(l,t)=|d(v,l)|-d(l,t)=d(v,l)-d(l,t)
\end{equation}
\begin{equation}
    \label{dual_alp_2}
    \pi_t^{DL}(v,2) = |d(v,l)-d(l,t)|-d(l,l) =|d(v,l)-d(l,t)| =|d(v,l)-d(l,t)|
\end{equation}
\begin{equation}
    \label{dual_alp_3}
    \pi_t^{DL}(v,3) = |d(l,l)-d(l,t)|-d(v,l) =|-d(l,t)|-d(v,l) =d(l,t)-d(v,l)
\end{equation}
Because we are taking the maximum, $\pi_t^{DL}(v,1)$ and $\pi_t^{DL}(v,3)$ simplify to the reverse triangle inequality.  $\pi_t^{DL}(v,4)$ and $\pi_t^{DL}(v,5)$ are, by their very definition, equal to the reverse triangle inequality, as well. $\pi_t^{DL}(v,6)$ cannot be used over the same set of landmarks because its equation would result in a division by zero. Therefore, this dual landmark ALP heuristic always reduces to a triangle inequality heuristic when the visited vertex and the target vertex share a landmark. Note that this does not mean that $\pi_t^{DL}$ necessarily reduces to the ALT heuristic in this case. For this to be the case, the shared landmark would have to be the landmark that would have maximized $\pi_t^{L}$. Also, we will see later, in Section \ref{sec:measuring}, that this ALP heuristic can still computationally outperform the ALT heuristic.
\par
This ALP heuristic function for dual landmarks is characterized in Section \ref{sec:analysis}. It should be noted that there are other quadrilateral inequalities for special cases and shapes that could also be used to define A* heuristics, as they, too, can yield estimates that never overestimate the shortest path. Quadrilaterals maintain these inequalities along with others that take into account perimeter, convexity, area, and a whole host of other geometric qualities that can be mapped to graphs. The examples for the dual landmark ALP heuristic used in this paper, however, are simply to demonstrate the ALP technique's utility in comparison to ALT.
\subsection{Distributed Embedding}
ALP can exhibit a space complexity of $\theta(|\textbf{\emph{L}}|^2 + |V|)$ using the following technique, called \emph{distributed embedding}. In the proposed dual landmark preprocessing for ALP, each landmark only computes the SPT for the subgraph induced by the landmark's graph partition. The only other calculation is an all-pairs shortest path calculation among the landmarks. For best results, each partition should be a connected subgraph to increase the likelihood that the shortest path from the landmark to any vertex in the partition lies entirely within the subgraph induced by the graph partition, though this is not a requirement.
\begin{figure*}
\centering
\includegraphics[scale=0.4]{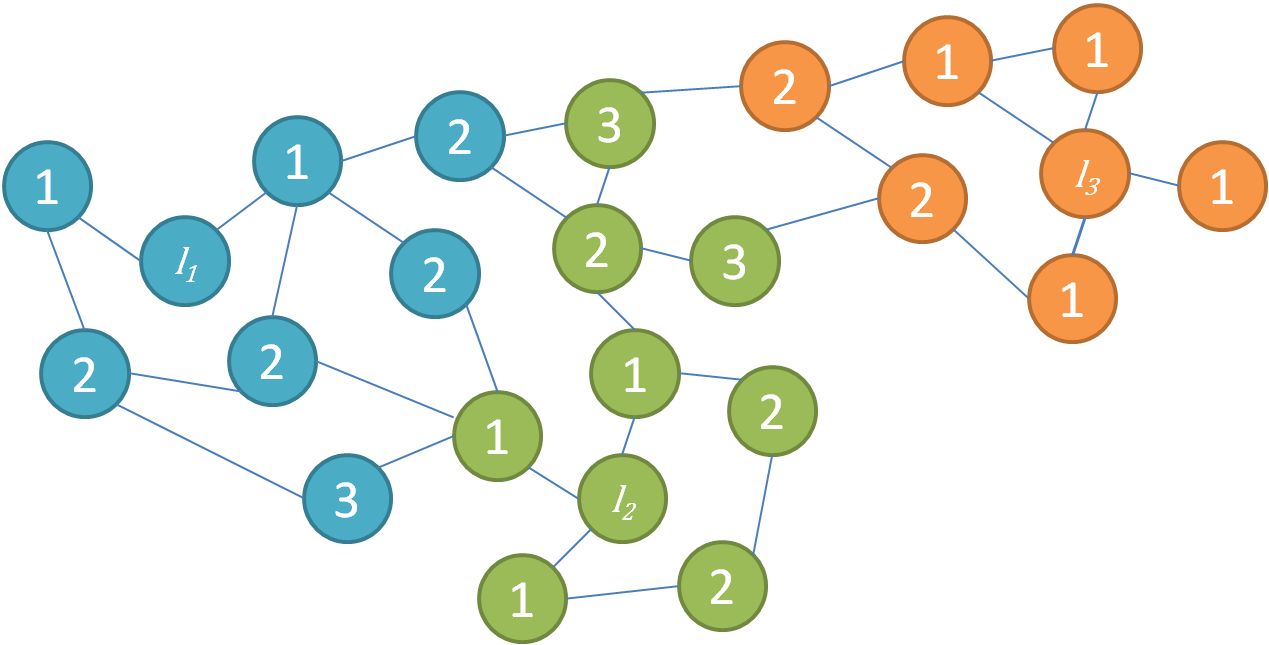}
\caption{Example of Vertex Labeling for the Dual Landmark ALP Heuristic with Distributed Embedding}
\label{fig:distributedembedding}
\end{figure*}
\par
An example of this distributed embedding technique is illustrated in Figure \ref{fig:distributedembedding}. During preprocessing, each vertex in the graph is labeled with an identifier, signifying its landmark partition and the distance to its corresponding landmark. Any of the embedding methods mentioned in Section \ref{subsec:embeddingmethods} can be used for the subgraph induced by the graph partition to select an optimal set. Once all landmarks have been chosen, an SPT for each landmark in $\textbf{\emph{L}}$ is then computed for its respective partition. This partitioning information is not explicitly stored by an external data structure. Rather, each vertex maintains distance information about the landmark to which it belongs along with a reference to that landmark. For landmark selection algorithms, if an algorithm requires understanding of all vertices that belong to a particular partition, then the partition can be discovered by finding all vertices with a common landmark reference. During query time, ALP carries out the normal A* algorithm with the ALP heuristic function, $\pi_t^{DL}$, that relies on polygon inequalities for quadrilaterals.
\par
Here, we describe distributed embedding in the context of the dual landmark approach, which leverages quadrilateral inequalities in the graph. This technique can identify other shapes to draw inequalities from (pentagons, hexagons, heptagons, etc.) by identifying multiple landmarks within a graph partition. For example, if two landmarks are chosen within each subgraph, the data structure contains enough information to leverage geometric inequalities for polygons with up to nine sides (nonagons). Choosing three landmarks within a subgraph would enable the leveraging of inequalities for polygons of up to sixteen sides (hexadecagons). The available inequalities are simply limited to the size of the graph (See full paper for details on how to compute the available inequalities for an embedding).
\section{Analysis of Dual Landmark Heuristics}
\label{sec:analysis}
In this section, we provide analysis for the dual landmark heuristic, $\pi_t^{DL}$, with respect to A* and ALT. For a source and target vertex pair, the following theorems apply to $\pi_t^{DL}$:
\begin{theorem}[Admissibility]
$\pi_t^{DL}$ is an admissible heuristic.
\begin{proof}
The proofs for the inequalities used for the heuristic are all derived in the previous section. Because the heuristic function has an upper bound set at the actual shortest path to the target, the heuristic will never overestimate the distance to the target, rendering it admissible.
\end{proof}
\label{theorem1}
\end{theorem}
\begin{theorem}[Non-monotonicity]
Using distributed embedding, $\pi_t^{DL}$ is not consistent.
\begin{proof}
Let $c$ be the cost of transitioning with A* from vertex $v$ to $v'$, for $v,v' \in V$. Recall that $c$ is nonnegative for the A* algorithm. Let $\pi_t^{DL}(v,1)$ be the maximum chosen for $\pi_t^{DL}$ for both of these iterations. Then, for $\pi_t^{DL}$ to be consistent,
\begin{equation}
    \label{consistency_proof2}
    |d(v,l_1 )-d(l_1,l_2 )|-d(t,l_2 ) \le c + |d(v',l_1 )-d(l_1,l_2 )|-d(t,l_2)
\end{equation}
Because $c$ is non-negative and the heuristic takes into account whether or not it moves towards or away from its landmark, $d(v',l_1 )=d(v,l_1 )-c$ or $d(v',l_1 )=d(v,l_1 )+c$, respectively. Therefore, this equation holds and demonstrates monotonicity over the same set of landmarks for successive iterations. However, allow the selection of landmarks for a query to change during the query, due to distributed embedding. For the heuristic to be consistent, with vertex $v$ belonging to landmark $l_i$ and $v'$ belonging to a different landmark $l_j$, once again let $\pi_t^{DL}(v,1)$ be the maximum chosen for $\pi_t^{DL}$ for both of these iterations. The following equation must then hold for $\pi_t^{DL}$ to be consistent:
\begin{equation}
    \label{consistency_proof3}
    |d(v,l_i )-d(l_i,l_2 )|-d(t,l_2 ) \le c + |d(v',l_j )-d(l_j,l_2 )|-d(t,l_2 )
\end{equation}
Let $l_j$ be chosen such that the $d(l_i,l_2 )>c+d(l_j,l_2)$ and $d(v,l_i )<d(v',l_j)$. This scenario yields a contradiction for the above equation. Therefore, $\pi_t^{DL}$ is not consistent.
\end{proof}
\label{theorem2}
\end{theorem}
\begin{theorem}
$\pi_t^{DL}$ does not dominate $\pi_t^{L}$ over the same set of landmarks.
\begin{proof}
In the previous section, we demonstrated that the dual landmark heuristic reduces to the triangle inequality heuristic over the same set of landmarks. The ALP estimate can then be, at most, equal to the estimate of ALT.
\end{proof}
\label{theorem3}
\end{theorem}
This means that, at best, the ALP heuristic estimates will be equal to that of ALT's and at worst, will always be less than ALT's.
\begin{theorem}
$\pi_t^{L}$ does not dominate $\pi_t^{DL}$ over an unequal set of landmarks.
\begin{proof}
This can be proven by contradiction. Let $\pi_t^{DL}(v,1)$ be the maximum chosen value for $\pi_t^{DL}$. For the triangle inequality heuristic to dominate the dual landmark heuristic:
\begin{equation}
    \label{dominate_proof}
    |d(v,l)-d(t,l)| \ge |d(v,l_i)-d(l_i,l_j)|-d(t,l_j)
\end{equation}
There are many scenarios that can contradict this statement. One example is when both $v$ and $t$ have equal distance from the ALT landmark and $d(l_i,l_j) > d(v,l_i) + d(t,l_j)$. In this case, the statement does not hold, yielding a contradiction.
\end{proof}
\label{theorem4}
\end{theorem}
To summarize, according to Theorem \ref{theorem1}, ALP's dual landmark heuristic is an admissible heuristic, making it a viable candidate for the A* algorithm, even though it is not consistent when using distributed embedding, as shown in the proof of Theorem \ref{theorem2}. From Theorem \ref{theorem3}, this heuristic for ALP does not dominate the heuristic for ALT over the same set of landmarks. From Theorem \ref{theorem4}, it is demonstrated that there are scenarios in which the ALP heuristic gives a higher estimation than the ALT algorithm. In the proof for Theorem \ref{theorem4}, a possible scenario for ALT (with all landmarks being equidistant to $v$ and $t$ such that the heuristic estimate is equal to 0) is used to theoretically demonstrate that it can have a lower heuristic estimate than ALP. The proof inherently shows the reverse, as well: that ALP can have a lower heuristic estimate than ALT.
\subsection{Comparing Landmark Configurations}
Recall that one A* heuristic outperforms the other, in terms of the number of vertices that are searched, by creating a higher estimation of the shortest path lower bound. In this section, we use this metric for performance to compare ALP's dual landmark heuristic to the ALT heuristic. Let $l_{\alpha} \in \textbf{\emph{L}}$ be the landmark chosen for ALT that maximizes its heuristic and $l_1,l_2 \in \textbf{\emph{L}}$ be the landmarks for the current vertex and the target, respectively.
\par
\textbf{Scenario 1}: $l_1 = l_\alpha \neq l_2$\\
Outperforms ALT when $|d(l_1,l_2 )-d(t,l_2 )|-d(v,l_1 ) \ge |d(v,l_1 )-d(t,l_1 )|$.\\
This scenario, in particular, outperforms ALT at the beginning of a search in a large graph, for $\pi_t^{DL}(v,2)$, \textbf{when the distances between the two landmarks is significantly large.} Particularly, if $|d(l_1,l_2 )-d(t,l_2)| \gg d(t,l_1) > d(v,l_1)$, the ALP heuristic will provide higher estimates than ALT. As such, $\pi_t^{DL}(v,3)$ and $\pi_t^{DL}(v,6)$ are the estimates that have a higher likelihood of yielding stronger results than the triangle inequality here.
\par
\textbf{Scenario 2}: $l_1 \ne l_\alpha =l_2$\\
Outperforms ALT when  $|d(l_1,l_2 )-d(v,l_1 )|-d(t,l_2 ) \ge |d(v,l_2 )-d(t,l_2)|$.\\
Particularly, if $|d(l_1,l_2)-d(v,l_1)| \gg d(v,l_2) > d(t,l_2)$, the heuristic dominates. Since we cannot rely on $d(v,l_1)$ to always be significantly larger than $d(v,l_2)$, the heuristic relies on the distance between the respective landmarks being significantly large to dominate. Therefore, in this scenario, the ALP heuristic dominates ALT \textbf{when the distance between the two landmarks is significantly large.} As such, $\pi_t^{DL}(v,1)$ and $\pi_t^{DL}(v,6)$ are the estimates that have a higher likelihood of yielding stronger results than the triangle inequality here.
\par
\textbf{Scenario 3}: $l_1 =l_\alpha =l_2$\\
Always has the same performance as ALT. $d(l_\alpha,l_\alpha)=0$, by definition. Therefore, all of the possible equations for the ALP heuristic are reduced to the triangle inequality. And \textbf{the ALP heuristic becomes the ALT heuristic.}
\par
\textbf{Scenario 4}: $l_1=l_2 \ne l_\alpha$\\
Outperforms ALT when  $|d(v,l_1 )-d(l_1,t)|-d(l_1,l_1) \ge |d(v,l_\alpha)-d(t,l_\alpha)|$.\\
$d(l_1,l_1 )=0$, by definition. Therefore, $\pi_t^{DL}(v,6)$ is eliminated as an option for the dual landmark heuristic. Because this occurs and because the ALT heuristic chooses the landmark that maximizes the triangle inequality, the best we can hope for is that the ALP heuristic is reduced to the heuristic for ALT. Therefore, \textbf{when the ALP algorithm's search is in the same partition, the ALP algorithm never dominates the ALT algorithm.}
\par
\textbf{Scenario 5}: $l_1 \ne l_\alpha \ne l_2$\\
Outperforms ALT when\\
$\pi_t^{DL}(v,1) \ge |d(v,l_\alpha )-d(t,l_\alpha )|$ or $\pi_t^{DL}(v,2) \ge |d(v,l_\alpha )-d(t,l_\alpha )|$ or\\
$\pi_t^{DL}(v,3) \ge |d(v,l_\alpha )-d(t,l_\alpha )|$ or $\pi_t^{DL}(v,6) \ge |d(v,l_\alpha )-d(t,l_\alpha )|$.\\
$\pi_t^{DL}(v,4)$ or $\pi_t^{DL}(v,5)$ can only reach the equivalence of the ALT heuristic's estimate over the same set of landmarks or for landmarks with similar distances to the one's used in ALT. Scenario 5 is the most common situational scenario. This is also the scenario that most significantly demonstrates that when the landmarks for ALT and ALP differ, the heuristic value for ALP is not always greater than the heuristic value for ALT, producing the results of Theorems 3 and 4.
\section{Pathfinding Performance}
\subsection{Measuring A* Heuristic Performance}
\label{sec:measuring}
Thus far, when describing ALP's performance in comparison to ALT, performance has been measured by the value calculated by a heuristic function. For A*, this value determines the size of the search space for any given query. However, one thing that is not taken into account in this and the proposal of other shortest path heuristics is the amount of processing needed to compute the actual heuristic at each visited vertex. In ALT, for each PPSP query, at each vertex, a number of subtractions equal to the number of landmarks is performed as well as a max operation. This means a $\theta(|\textbf{\emph{L}}|)$ runtime for every visited node. For large-scale graphs, which require more landmarks to be preprocessed, this can significantly add to the overall compute time of queries. In comparison, with the proposed dual landmark ALP heuristic, if the visited vertex and the target vertex are owned by different landmarks, exactly nine subtraction operations, two multiplication operations, and a single division operation occurs with an $O(4)$ max operation. Over the same set of landmarks, dual landmark ALP executes eight subtraction operations and an $O(5)$ max operation upon visiting a vertex. This means that, in terms of practical, processor-based performance measurements, even over the same set of landmarks, it is possible for dual landmark ALP to outperform ALT. Because Dijkstra's algorithm performs no arithmetic operations at each visited vertex, it is possible for Dijkstra's algorithm to outperform both ALT and ALP algorithms at large scales (See Section \ref{sec:measuring} in full paper).
\subsection{Computational Complexity}
The benefits in time and space complexity of ALP over ALT is due to the novelty of distributed embedding. Note that the worst-case time complexity for preprocessing of ALP remains the same as that of ALT. The actual shortest path between two vertices within a graph partition could include vertices from outside the partition. This means that, in the worst case, the generated SPT for a landmark within a subgraph includes the entire vertex set of the graph. This phenomena would rarely happen in practice. In practice, the SPT will be significantly small in comparison to the size of the graph and its generation will run in a fraction of the time. Therefore, for a graph in which the vertices of each partition match the vertices in a partition's shortest path tree, let $E'$  be the average number of edges in each partition and V' the average number of vertices in each partition. Then the average runtime of ALP preprocessing, not including landmark selection, is
\begin{equation}
\theta (|\textbf{\emph{L}}| \cdot (|E'|+|V'|  \log{|V'|}))
\end{equation}
Because the shortest path tree is computed from every chosen landmark and distance along with an all-pairs shortest path calculation for the landmarks, ALT's data structure requires $\theta(|\textbf{\emph{L}}| \cdot |V|+|\textbf{\emph{L}}|^2)$ space, where $\textbf{\emph{L}}$ is the set of chosen landmarks. Since $|\textbf{\emph{L}}| \le |V|$, the theoretical space requirement for ALT can be said to be $O(|V|^2)$. Note that this upper limit is only theoretical, as a relatively small number of landmarks are chosen for any particular graph. Therefore, the $\theta(|\textbf{\emph{L}}| \cdot |V|+|\textbf{\emph{L}}|^2)$ space requirement is a more practical specification. For ALP, shortest path data is stored for the landmark-vertex pairs of each graph partition and the pairwise distances between landmarks. Therefore, ALP's data structure requires $\theta(|V|+|\textbf{\emph{L}}|^2)$ space. Once again, because $|\textbf{\emph{L}}| \le |V|$, the space requirement for ALP can also be described as $O(|V|+|V|^2 )=O(|V|^2)$, which is theoretically larger than the worst-case ALT requirement.
\section{Conclusions}
In this paper, we identify a heuristic for A* that leverages a data structure of size $\theta(|\textbf{\emph{L}}|^{2} + |V|)$ as opposed to ALT's previous $\theta(|\textbf{\emph{L}}|\cdot|V|)$. This data structure is formed through a new embedding process, which requires identifying distances only within a graph partition. We've also identified each theoretical scenario in which ALP's estimates are greater than ALT's. Finally, we have established that in cases in which the ALT heuristic has greater average estimates than the ALP dual landmark heuristic, ALP can still computationally outperform ALT and Dijkstra's algorithm can outperform A* using either preprocessing technique. The fact that Dijkstra's algorithm can outperform both of these methods as graphs scale should serve as a cautionary example for other methods of shortest path preprocessing.
\section{Acknowledgments}
Special thanks to advisor Dr. Michael J. Laszlo and my doctoral committee at the Nova Southeastern University GSCIS for their support throughout the Computer Science PhD program. Also, special thanks to my employer, Raytheon BBN Technologies, for mentorship, guidance, and financial aid throughout the course of the CISD program.
\bibliographystyle{abbrv}
\bibliography{refs}
\end{document}